\documentclass[prd, nofootinbib, floatfix]{revtex4}

\usepackage{amsmath, natbib}
\usepackage{amsbsy}

\input epsf
\usepackage{amsmath,amssymb,subfigure}
\usepackage{graphicx}
\usepackage{epsfig}
\usepackage{color}

\begin{document} 

\title{Classification of Stellar Spectra with LLE} 
\author{Scott F.\ Daniel, Andrew Connolly, Jeff Schneider, Jake Vanderplas, and
Liang Xiong}

%\affiliation{$^1$Institute for the Early Universe, Ewha Womans University, 
%Seoul, Korea\\ 
%$^2$Lawrence Berkeley National Laboratory, Berkeley, CA, USA\\ 
%$^3$Berkeley Center for Cosmological Physics, University of California, 
%Berkeley, CA, USA 
%}

\date{\today}

%%%%%%%%%%%%%%%%%%%%%%%%%%%%%%%%%%%%%%%%%%%%%%%%%%%%%%%%%%%%
%%%%%%%%%%%%%%%%%%%%%%%%%%%%%%%%%%%%%%%%%%%%%%%%%%%%%%%%%%%%
\begin{abstract} 
  We investigate the use of dimensionality reduction techniques for
  the classification of stellar spectra selected from the SDSS. Using
  local linear embedding (LLE), a technique that preserves the local
  (and possibly non-linear) structure within high dimensional data
  sets, we show that the majority of stellar spectra can be
  represented as a one dimensional sequence within a three dimensional
  space. The position along this sequence is highly correlated with
  spectral temperature. Deviations from this ``stellar locus'' are
  indicative of spectra with strong emission lines (including
  misclassified galaxies) or broad absorption lines (e.g.\ Carbon
  stars). Based on this analysis, we propose a hierarchical
  classification scheme using LLE that progressively identifies and
  classifies stellar spectra in a manner that requires no feature
  extraction and that can reproduce the classic MK classifications to
  an accuracy of one type. %this is the worst error in Fig 10
\end{abstract} 

\maketitle

%%%%%%%%%%%%%%%%%%%%%%%%%%%%%%%%%%%%%%%%%%%%%%%%%%%%%%%%%%%%
%%%%%%%%%%%%%%%%%%%%%%%%%%%%%%%%%%%%%%%%%%%%%%%%%%%%%%%%%%%%
\section{Introduction \label{sec:intro}}

Automatic classification of stellar data is a problem as old as the
use of computers in astronomy.  Since survey projects have begun
presenting us with spectral data from literally hundreds of thousands
of sources, it has become untenable to classify all of them ``by
hand.''  Computer science provides several tools and algorithms
available to help us alleviate the human experts' work load.  Neural
networks \cite{StorrieLombardi:1994,Singh:1998} and Principal
Component Analysis (PCA; Deeming 1964) are among the most popular
computational tools presently used to tackle large sets of
astronomical data.  In this paper, we will consider a relatively new
method: Local Linear Embedding (LLE) \cite{Roweis:2000}.

At the heart of automated stellar classification is dimensionality
reduction: taking a large number $N$ of $D\gg 1$ dimensional data (in
this paper we consider $N=49,529$ stellar spectra sampled over $D=500$
wavelength bins), and projecting the data onto a basis such that the
first $d\ll D$ dimensions contain the bulk of the physical information
encoded in the data.  LLE attempts to reduce the dimensionality of the
input data points while preserving the non-linear relationships
between them.  It does so by analyzing the data incrementally, in
small neighborhoods, rather than all at once.  This is particularly
useful for astronomical classification, as we shall see below, since
it can be simultaneously sensitive to continuum and line shapes.
Thus, LLE ought to provide a more robust object classification from
fewer projected dimensions than PCA.  Vanderplas and Connolly (2009)
explored LLE as a means of characterizing the spectral energy
distributions of galaxies.  They found the method very effective and,
in some cases, more accurate than traditional tests at distinguishing
different types of galaxies (broad- and narrow-line QSO's, emission line
galaxies, quiescent galaxies, and absorption galaxies; see their
Figure 2) without the need to identify and measure individual features
in the galaxies' spectra.  We use their code\footnote{Publically
  available at http://ssg.astro.washington.edu/software} to analyze
49,529 stellar spectra from the SDSS Data Release 7
(DR7)~\cite{Abazajian:2008wr}.  We specifically consider the ability
of LLE to identify different types of objects (galaxies, stars, and
QSOs).  In the case of stars, we consider the ability of LLE
to classify objects according to
their MK spectral types.  In Section \ref{sec:PCA} we will discuss
past work on automated stellar classification using Principal
Component Analyis (PCA).  In Section \ref{sec:LLE} we will review the
algorithm of LLE and contrast it with that of PCA.  In Section
\ref{sec:results} we will present the results of spectral
classification using only LLE analysis.  In Section
\ref{sec:discussion} we will contrast our results with those from PCA.

%%%%%%%%%%%%%%%%%%%%%%%%%%%%%%%%%%%%%%%%%%%%%%%%%%
\section{Past Work with Principal Component Analysis \label{sec:PCA}}

Most of the work to date has focused on Principal Component Analysis
(PCA).  PCA takes the data vectors and projects them onto an
orthogonal basis made up of the eigenvectors of the correlation matrix
of the unprocessed data.  Theoretically, the originally
high-dimensional data can be well-characterized by linear combinations
of the few eigenvectors with the highest
eigenvalues~\cite{BailerJones,McGurk:2010aw}.  The use of PCA in
analyzing large volumes of stellar data was first proposed by Deeming
(1964).  Deeming considered data from measurements of five spectral
lines from G and K giants, projected them using PCA, and found that
the component of each vector in the first projected coordinate was
sufficient to characterize the corresponding star's MK spectral type.
This is obviously an oversimplified case since it deals with a data
set comprised entirely of late-type stars.  Figure 4 of Singh (2001)
shows that adding early-type stars to the data set necessitates the
addition of at least one extra PCA dimension to recreate Deeming's
successful classification.
  
Following Deeming, much of the work in PCA and stellar classification
focused on using PCA to pre-process the data fed into artificial
neural networks.  Storrie-Lombardi {\it et al}. (1994) found that data
must be projected onto a minimum of three eigencomponents in order to
improve the MK classification performance of raw-data neural networks.
Singh {\it et al}. (1998) found that at least 10 components were
necessary to accurately reproduce classifications performed by human
experts, but that use of between two and five eigencomponents
reproduced human results to within three spectral subtypes (see their
Table 2).  This is the predominant theme of stellar classification
with PCA projection: at least two eigencomponents are needed to
reproduce the results of human experts
\cite{Christian:1982,Whitney:1983a,Beauchemin:1991,Singh:2001}.  We
hope to show that LLE can improve on this performance by producing
a single parameter that uniquely correlated with spectral type.

In addition to reproducing the MK classification of stellar spectra,
there has been significant interest in using PCA to extract the
physical properties (effective temperature, metallicity, gravity, {\it
  etc}.) of stars \cite{Whitney:1983b,McGurk:2010aw}.  Using stellar
spectra from the SDSS DR7 \cite{Abazajian:2008wr}, McGurk {\it et al.}
(2010) found a correlation between a star's PCA decomposition and its
metallicity.  We will also consider the ability of LLE to extract
physical information from stellar spectra.

Recently, PCA has also been applied to classifying the spectral energy
distribution of galaxies
\cite{Connolly:1994ph,BailerJones,Ronen:1999,Yip:2004a}, and quasars
\cite{Yip:2004b}.  Cabanac {\it et al}. (2002) found that PCA can be
effective at separating out stellar, galactic, and QSO spectra while
simultaneously dividing the stellar spectra into spectral classes.
Because our initial filtering of the data does not successfully remove
all non-stellar spectra from our data set, we are able to test the
ability of LLE to separate galactic from stellar sources as well.

%%%%%%%%%%%%%%%%%%%%%%%%%%%%%%%%%%%%%%%%%%%%%%%%%%
\section{The LLE Algorithm \label{sec:LLE}}

One shortcoming of PCA (already identified by Deeming in 1964) 
is that it relies on the assumption that all of
the data considered can be well-described as a linear combination of
all of the other data.  If the data instead conforms to 
some underlying non-linear
manifold (for example, if not only the presence but the 
intensity of lines varies between spectra; Connolly {\it et al.} 1995), 
PCA will wash that information out. 
LLE avoids such oversimplification
by attempting to reconstruct each data point from a linear combination of only
its $k$ nearest neighbors, where $k\ll N$.  This local reconstruction is characterized in terms of a global weight
matrix ${\bf W}$, i.e. the $i$th datapoint $\vec{x}^i$ is reconstructed as
\begin{equation}
\label{eq:weight}
\vec{x}^i=\sum_j^{N} {\bf W}_{ij}\vec{x}^j
\end{equation}
where ${\bf W}_{ij}=0$ if the $j\text{th}$ 
data point is not one of $\vec{x}^i$'s $k$
nearest neighbors.  
Solving for ${\bf W}_{ij}$ is later referred to as the ``training phase.''  
LLE projects the data down to a $d\ll D$-dimensional space by
finding $d$-dimensional data points $\vec{y}$ 
such that the relationship (\ref{eq:weight})
remains true, i.e.,
\begin{equation}
\vec{y}^i=\sum_j^N{\bf W}_{ij}\vec{y}^j
\end{equation}  
for the same weight matrix as above.
The relationships between data points and their nearest
neighbors remain linear, but the non-linear relationships between
disparate neighborhoods are preserved.  
This final projection is an eigenvalue problem such that the
$d$ projected dimensions correspond to the eigenvectors of the matrix
\begin{equation}
\label{eq:mmatrix}
{\bf M}=({\bf I}-{\bf W})({\bf I}-{\bf W})^T
\end{equation}
with the smallest eigenvalues\footnote{The smallest eigenvalue actually
corresponds to a global translation of the data and is discarded.}.
For a more detailed description see Roweis and Saul (2000), de Ridder and Duin
(2002) and Vanderplas and Connolly (2009).

\begin{figure}[!t]
\subfigure[]{\includegraphics[scale=0.5]{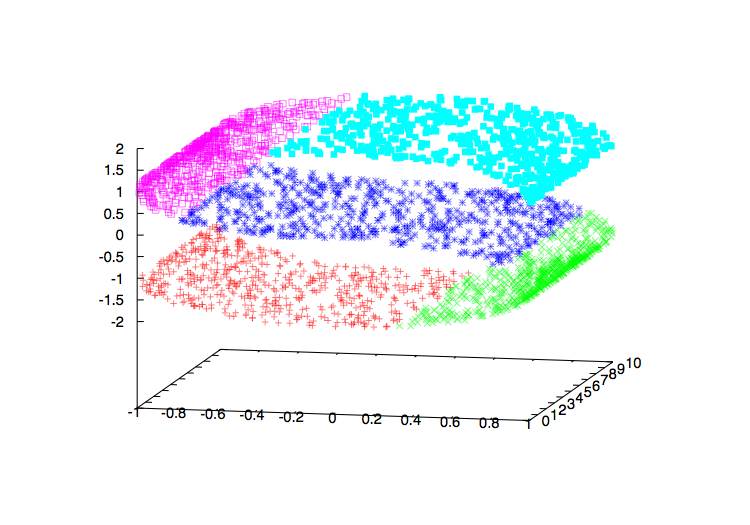}
\label{fig:s_plot}}
\subfigure[]{\includegraphics[scale=0.5]{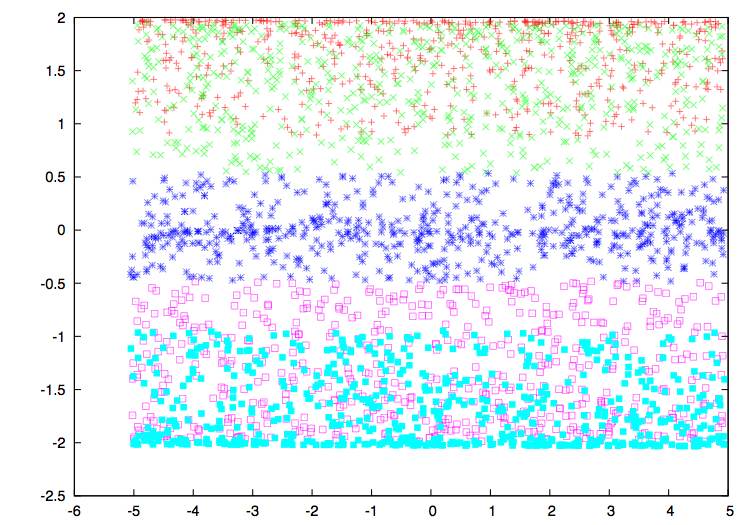}
\label{fig:pca_plot}}
\subfigure[]{\includegraphics[scale=0.5]{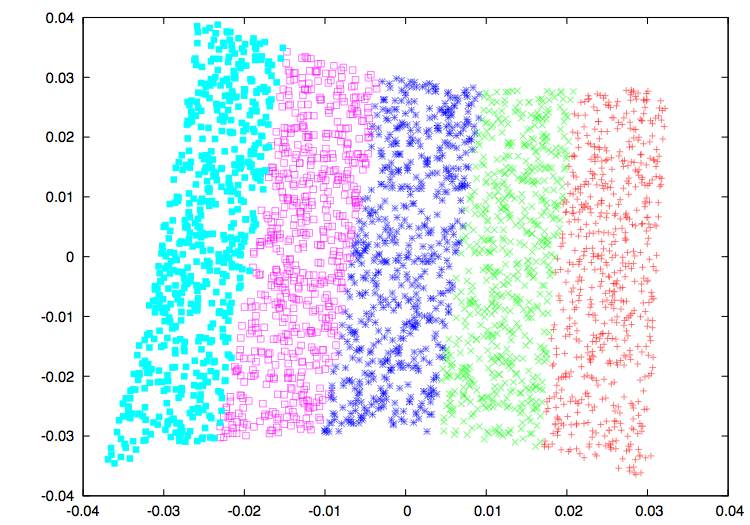}
\label{fig:flag_plot}}
\caption{
A demonstration of the advantage of LLE over PCA.  Figure \ref{fig:s_plot} shows
the unprocessed, 
three-dimensional input data.  Figure \ref{fig:pca_plot} shows
the result of a simple, two-dimensional PCA projection of the data.  Figure
\ref{fig:flag_plot} shows the result of a two-dimensional LLE projection of the
data.  Color-coding is consistent between samples.  Note that the PCA projection
confuses the relationship between points where curvature in Figure
\ref{fig:s_plot} is strongest.  LLE correctly maps the data to its underlying
manifold.
}
\label{fig:example}%
\end{figure}

Figure \ref{fig:example} demonstrates the advantage of this
neighborhood-by-neighborhood treatment of the data.  Figure \ref{fig:s_plot}
posits some three-dimensional data on an embedded two-dimensional manifold. 
Figure \ref{fig:pca_plot} attempts to unwind the manifold using a simple
two-dimensional PCA projection as described in Section 3 of Xiong et al (2011).  
The color bands are correctly ordered, however, they blend together
where the embedding goes non-linear (where the `S' shape curves in Figure
\ref{fig:s_plot}).  Figure \ref{fig:flag_plot} attempts to unwind the manifold
using LLE.  There is no blending.  Because it divides the data into small
neighborhoods, LLE is always treating quasi-linear subsets of the entire
manifold.  Linear projection as in equation (\ref{eq:weight}) remains valid and
the correct relationship between datapoints is preserved.  This will be
especially useful in spectroscopy, where objects can be related by their
continua, their emission and absorption features, or some non-trivial
combination of the two.

One disadvantage of the LLE algorithm is that, unlike PCA, it does not 
produce basis vectors onto which future data points can be projected for
decomposition.  The weight matrix in equation (\ref{eq:weight}) means that the
projection is dependent upon and unique to the points being projected.  
To add new points to a data set, one must redo the entire analysis from scratch.
Conversely, in PCA, one simply decomposes the new points into 
the basis vectors already solved for.
There are
ways to modify the algorithm to avoid this complication.  
In their section 5.3, Vanderplas and Connolly (2009)
propose a method in which, for each of the $N$ data points, 
the neighbors used to construct 
the weight matrix are chosen from only a subset of the data.  This subset
is selected to maximize the represented signal variance relative to
the full data set.  When new data points are added under this modification, one
only needs to perform the nearest-neighbor search for the new points (as opposed
to the full data set in the context of the new points) and re-run the algorithm
from there.  We adopt this modification in the work below, training our LLE
projection on only 5,000 of the full 49,529 spectra in the data set.

Because LLE attempts to subdivide the original, non-linearly related 
data set into smaller, linearly-related data sets, it may not be applicable to
data sparsely sample the non-linear space.  It may be possible
to address this concern by reducing the number of nearest neighbors used in
constructing the weight matrix \ref{eq:weight}.  It would probably be safer to
avoid using LLE on small data sets.

%%%%%%%%%%%%%%%%%%%%%%%%%%%%%%%%%%%%%%%%%%%%%%%%%%%%%
\section{Results \label{sec:results}}

We use the same set of stellar spectra that Xiong {\it et al}. (2011) 
treat with PCA.  Objects are chosen from the 85,564 DR7
sources classified as science-quality stellar spectra (though, as will
be seen, further analysis reveals that some galactic spectra are
included).  Objects are rejected if they have redshift $z>0.36$, more
than 20\% of the pixels in the image are bad, their spectra contain
large positive ($>10^4$) or negative ($<-100$) spikes in flux, their
signal-to-noise ratio is less than 10, or their magnitude is less than
15.5.  This results in 49,529 spectra to analyze.  These spectra
are reduced to 500 wavelength bins evenly spaced in
$\log_{10}(\lambda)$.  Gaps in the spectra due to bad pixels are
filled in using the PCA eigenspectra of Xiong {\it et al} (2011).  We
transform the spectra to their rest-frame wavelength and normalize the flux
so that the total flux (including both
continua and lines) is the same for all spectra.
We perform our analysis using the LLE code made public by Vanderplas
and Connolly (2009).  We use $k=15$ neighbors in the training phase
and set the projected dimensionality so that $95\%$ of the variance in
the usual sample covariance matrix is preserved \cite{deRidder:2002}.
In practice, this means that we are projecting onto 10 dimensions,
though we find that only the first three are interesting for stellar
characterization.

\subsection{Gross Features of the LLE Projection \label{sec:gross}}
 
\begin{figure}[!t]
\subfigure[]{\includegraphics[scale=0.85]{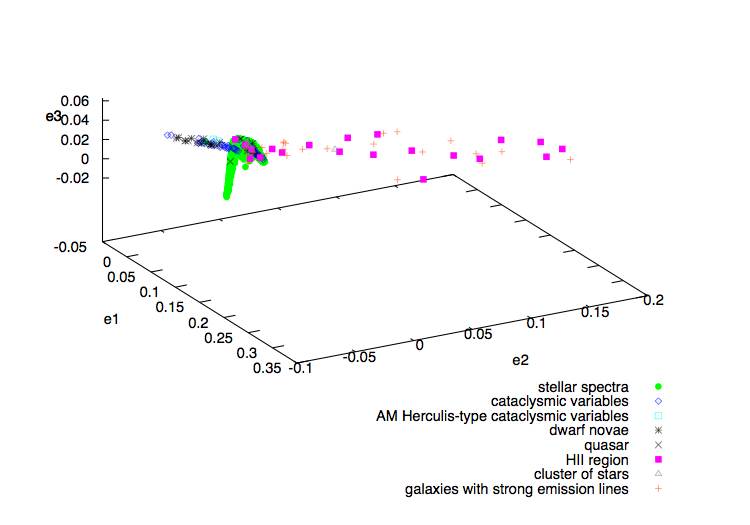}
\label{fig:boom_3d}}
\\\subfigure[]{\includegraphics[scale=0.85]{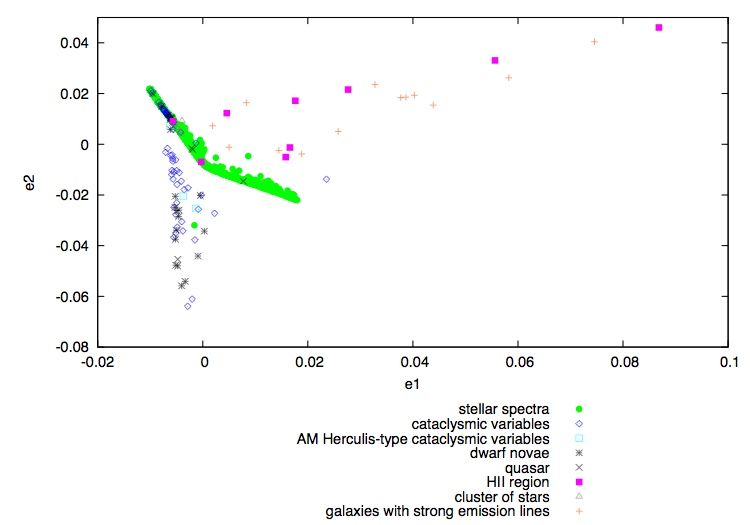}
\label{fig:boom_2d}}
\caption{
Figure \ref{fig:boom_3d} plots the SDSS stellar spectra in the
first three projected LLE dimensions.  Coordinates are named such
that $e1$ corresponds to the first coordinate of the LLE decomposition.
$e3$ corresponds to the third.  Object classifications in the legend
derive from the SIMBAD 
database (http://cds.u-strasbg.fr/cgi-bin/Otype?X).
Figure \ref{fig:boom_2d} shows the same plot in only the $e1$ and $e2$
coordinates and zoomed in to accentuate the structures
formed by emission line galaxies and cataclysmic variables.
}
\label{fig:boom}%
\end{figure}

Figures \ref{fig:boom} plot the projection of the data onto these
first three LLE dimensions.  These dimensions are labeled so that $e1$
is the dimension corresponding to the smallest retained eigenvalue of
the matrix (\ref{eq:mmatrix}), $e2$ is the dimension corresponding to
the second-smallest retained eigenvalue, and so on.  Spectra are
sorted according to classifications found by comparing DR7 to the
SIMBAD database\footnote{http://simbad.u-strasbg.fr/simbad/sim-fid}.

Most objects cluster in a quasi-parabolic feature in the $e1<0.05$,
$e2<0$ region of this coordinate space.  We will hereafter refer to
this feature as the ``stellar locus.''  The most remarkable feature of
this locus is that it appears as a one parameter sequence that is
consistent with the classification of spectra by temperature. Notable
exceptions are two independent branches: one formed mostly of galactic
emission-line spectra that accidentally escaped our initial filtering
of the data and one formed of Cataclysmic Variables and Dwarf Novae.

\begin{figure}[!t]
\subfigure[]
{\includegraphics[scale=0.3]{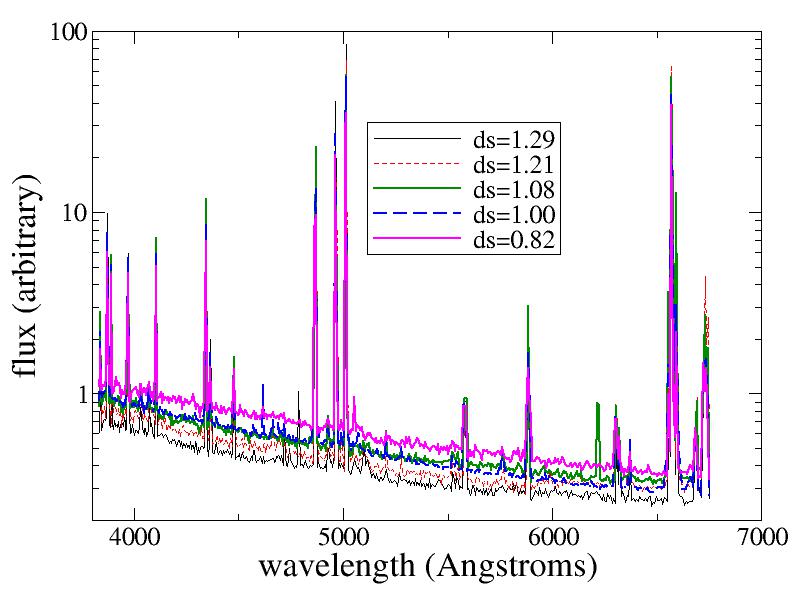}
\label{fig:gex}}
\subfigure[]
{\includegraphics[scale=0.3]{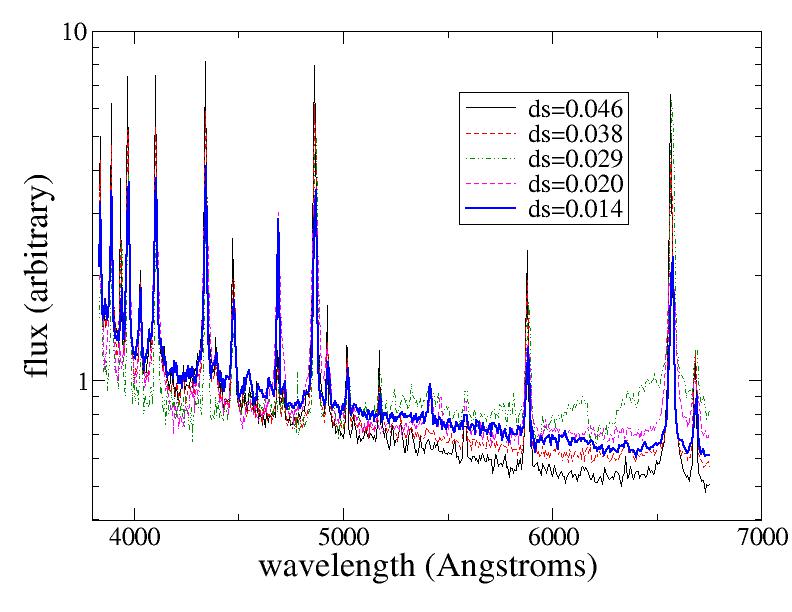}
\label{fig:cvex}}
\caption{
Figure \ref{fig:gex} plots example spectra from the galaxy branch in Figure
\ref{fig:boom}.  The quantity ds is the distance of the corresponding data point
from the stellar locus in $\{e1,e2,e3\}$ space.  
The flux axis is logarithmic to show both the extreme emission
features and the apparent dependence of ds on the continuum.
Figure \ref{fig:cvex} plots example spectra from the CV/DN branch in Figure
\ref{fig:boom}.
}
\label{fig:g_cv_branch}
\end{figure}

Figure \ref{fig:boom_2d} plots the LLE projection in only two
dimensions.  We see that the galaxy branch extends primarily in the
$e1$ direction, while the Cataclysmic Variable/Dwarf Nova 
branch is nearly parallel to the $e2$
direction.  This would seem to indicate that these two projected
coordinates correspond to features that dominate in these two classes
of objects.  Figures \ref{fig:gex} and \ref{fig:cvex} plot example
spectra taken from along these branches in LLE space.  In both cases,
the spectra are dominated by strong emission line features.  The
quantity ds labelling each of the spectra is their distance in LLE
space from the stellar locus.  Inspection of Figure \ref{fig:gex}
reveals that, along the galaxy branch, ds correlates with both the intensity 
in continuum and line emission.
%{\bf AJC: Is this correct - is it not
%  line strength. I thought we normalized out the overall continuum
%  emission;  SFD: we normalized the total flux, including lines.  The continua
%  can still differ after that}.  
In the case of the
CV/DN branch, it is not clear that there is any correlation between ds
and the continuum.  However, the strengths of many different emission
features do correlate with ds, as is shown in Figure \ref{fig:cvex}.

\begin{figure}[!t]
\includegraphics[width=\columnwidth]{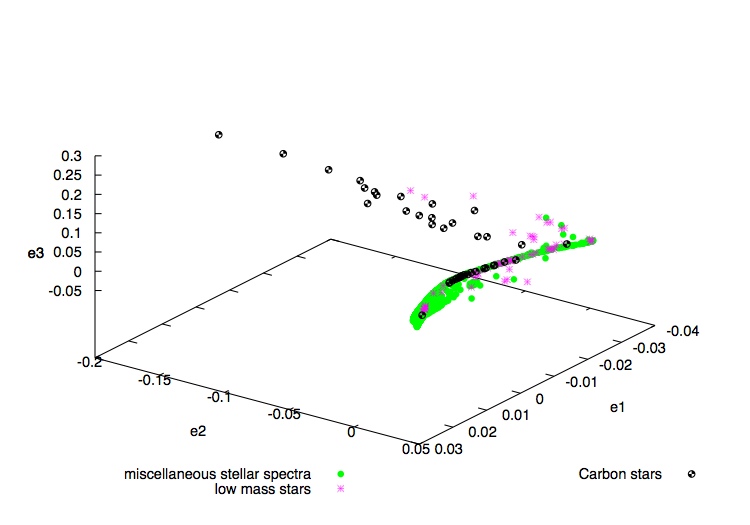}
\caption{
The first three LLE-projected dimensions of our data set with
those objects that exist in the emission line galaxy and CV/DN branches of 
Figure \ref{fig:boom} removed.  The stellar locus 
is now more closely confined to
the $e1$ and $e2$ dimensions.  Carbon stars extend into the $e3$ dimension.
}
\label{fig:justboom}%
\end{figure}

\begin{figure}[!t]
\subfigure[]{
\includegraphics[scale=0.3]{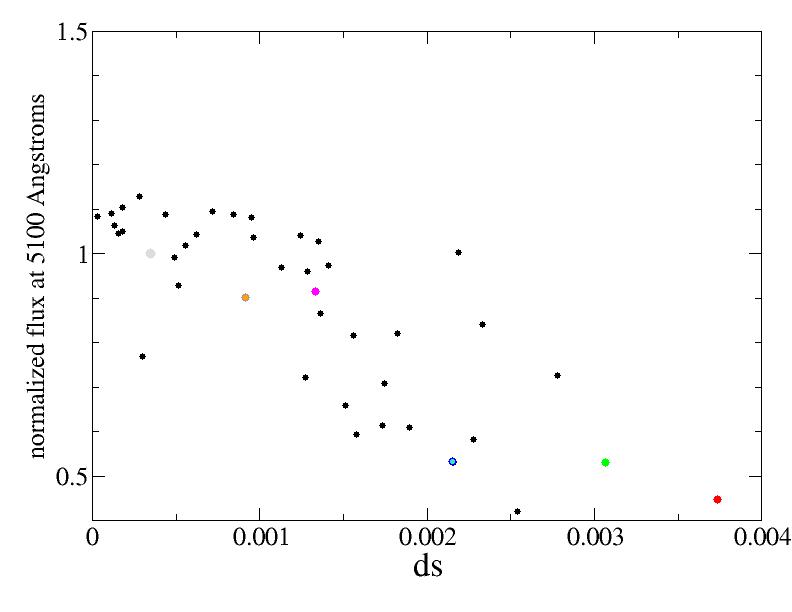}
\label{fig:cflux}
}
\subfigure[]{
\includegraphics[scale=0.3]{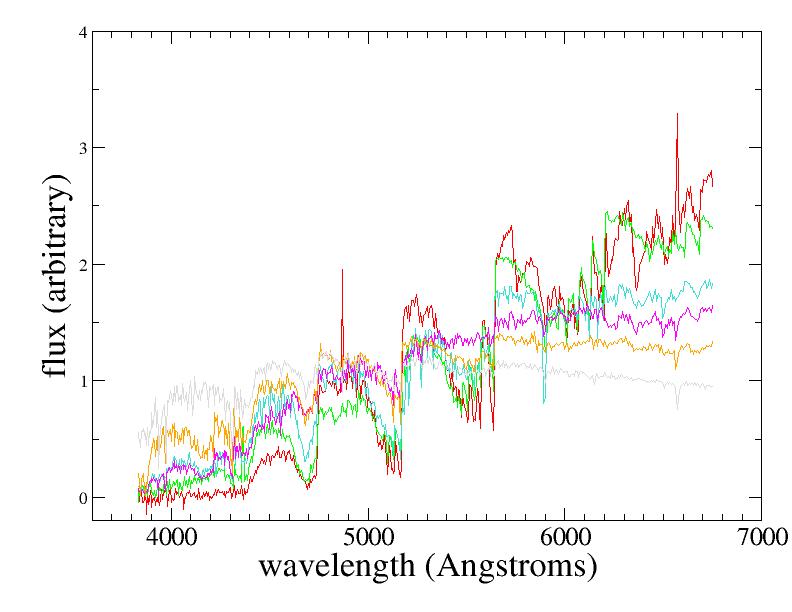}
\label{fig:cex}
}
\caption{
Figure
\ref{fig:cflux} plots flux in the one broad emission feature versus distance from the 
stellar locus in Figure \ref{fig:boom}.  
Colored points correspond to the example spectra plotted in Figure \ref{fig:cex}.
These figures show only spectra positively 
identified as carbon
stars.
}
\label{fig:carbonexample}%
\end{figure}

Removing the data points along these branches from the data set and
re-performing our analysis, we find that the same quasi-parabolic
stellar locus persists, but much more closely confined to a plane in
$e1$ and $e2$, while, in this new projection, carbon stars extend far
into the $e3$ direction.  We present this result in Figure
\ref{fig:justboom}.  Figure \ref{fig:cex} plots examples of
carbon star spectra.  They seem to be characterized according to the
presence or absence of broad absorption features, e.g. at $\sim 5100~\AA$.  
%This correlation is present even in our original projection.
%Figure \ref{fig:cflux} plots flux in this feature versus distance from the
%stellar locus in Figure \ref{fig:boom}.

There is, therefore, a natural progression in the LLE decompositions
of the stellar spectra. Deviations from a smooth continuum (e.g.\
strong emission lines or broad absorption features) perturb the
positions of sources away from the ``stellar locus'' and are
identified as outliers within the LLE projected space. Excluding
spectra with these emission or absorption features from the projection
enables spectra with weaker features to be identified until we are left
with a low dimensional series of continuum only stellar spectra. This
approach provides both a mechanism for identifying anomalous spectra
and for classifying the normal populations.

\begin{figure}[h]
\center
\includegraphics[width=\columnwidth]{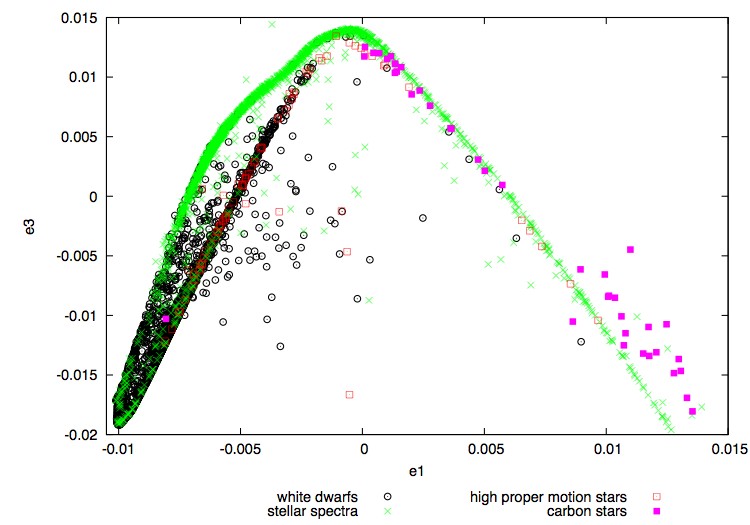}
\caption{
Figure \ref{fig:boom_3d} plotted in the $e1$ and $e3$ coordinates to highlight
the structure of the stellar locus.  The $e1<0$ half of the plot shows
a separate branch beneath the stellar locus.  This branch is made up
primarily of white dwarfs and
high proper motion stars.
}
\label{fig:wd}
\end{figure}

Figure \ref{fig:wd} plots the stellar locus of the original projection 
from Figure \ref{fig:boom} 
in just two dimensions ($e1$ and $e3$).  
Looking at the $e1>0$ leg of the stellar locus in
this plot, one sees that, even in the original LLE projection, Carbon stars
are significantly scattered away from the locus.
This is quantified in Figure \ref{fig:cflux}, where the quantity 
ds is an
object's distance from the stellar locus in Figure \ref{fig:boom}.  We again
see a correlation with flux in the feature at $\sim 5100~\AA$. We will revisit
this parallel between the projections in Figures \ref{fig:boom} and
\ref{fig:justboom} in Section \ref{sec:layers}.

\begin{figure}[!t]
\includegraphics[width=\columnwidth]{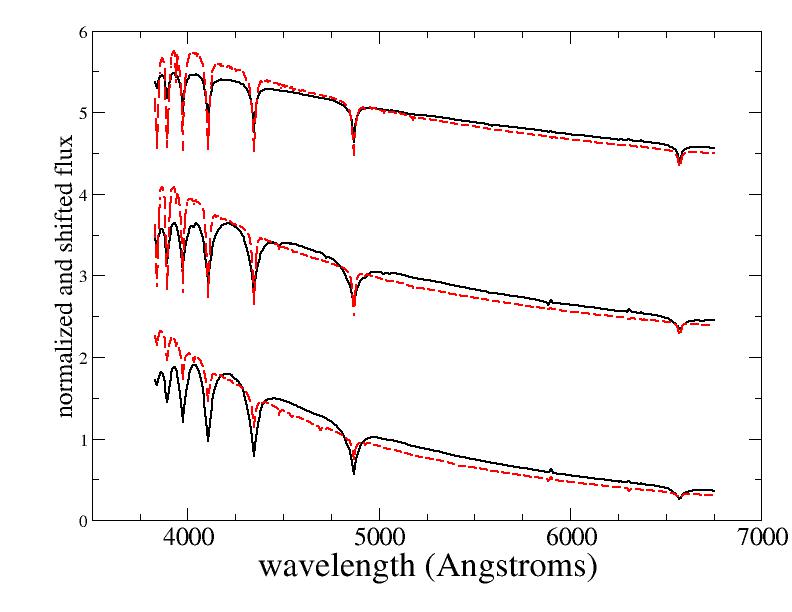}
\caption{
We compare average spectra along the $e1<0$ section of the 
stellar locus and the
separate White Dwarf branch in Figure \ref{fig:wd}.  Solid (black) curves
correspond to spectra that are closer to the White Dwarf branch than
the main stellar locus.  
Dashed (red) curves correspond to spectra that are
closer to the stellar locus 
than the White Dwarf branch.  From bottom to top, curves
correspond to averages about $e1=-0.01,-0.008,-0.006$.
}
\label{fig:wd_spectra}%
\end{figure}

LLE projection also produces some 
useful substructures within the stellar locus.
From the overhead perspective of Figure \ref{fig:boom_2d}, the
stellar locus looks effectively
one-dimensional.  
The same is evident in the two-dimensional projection of Figure \ref{fig:wd}.
The only exception is a bifurcation in the $e1<0$ region. 
Here, the White Dwarfs and high proper motion stars have separated
themselves out into their own branch independent (and below in this projection) 
of that occupied by other stellar classes.  In Figure
\ref{fig:wd_spectra}, we segregate the $e1<0$ data points 
from Figure \ref{fig:wd} into two groups: those closer to the White Dwarf
branch than the stellar locus, and those closer to the stellar locus.  We then
average the spectra with similar $e1$-coordinate values and plot them on top of
each other.  We find that the principal difference between 
spectra in the White
Dwarf branch and spectra in the stellar locus occurs in the continuum flux at short
wavelengths.  This is consistent with our supposition that the stellar locus
is a continuum-dominated feature of our projection.

\subsection{Spectral Classification Using LLE \label{sec:classification}}

\begin{figure}[!t]
\center
\subfigure[]{
\includegraphics[scale=0.5]{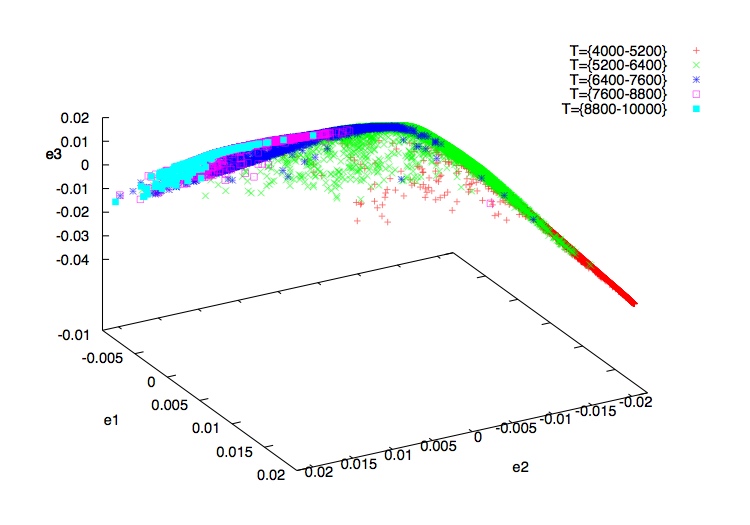}
\label{fig:temp}
}
\subfigure[]{
\includegraphics[scale=0.5]{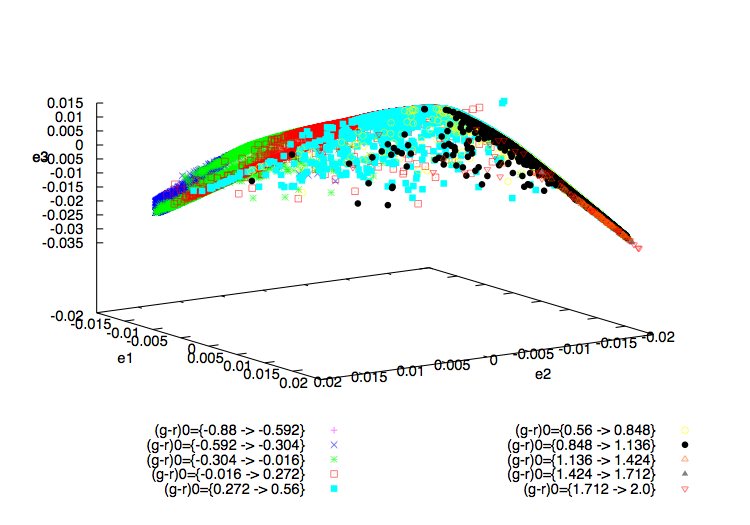}
\label{fig:color}
}
\caption{
The same as Figure \ref{fig:boom_3d} except that now objects are classified
according to effective temperature (\ref{fig:temp})
and $(g-r)_0$ color (\ref{fig:color}).  
As you can see, one-dimensional position
in the stellar locus  correlates well with effective temperature 
(there are no available effective temperatures or colors for objects
far from the stellar locus).
}
\label{fig:tempandcolor}
\end{figure}

\begin{figure}[!t]
\subfigure[$4000<T_\text{eff}<5200$]
{\includegraphics[scale=0.3]{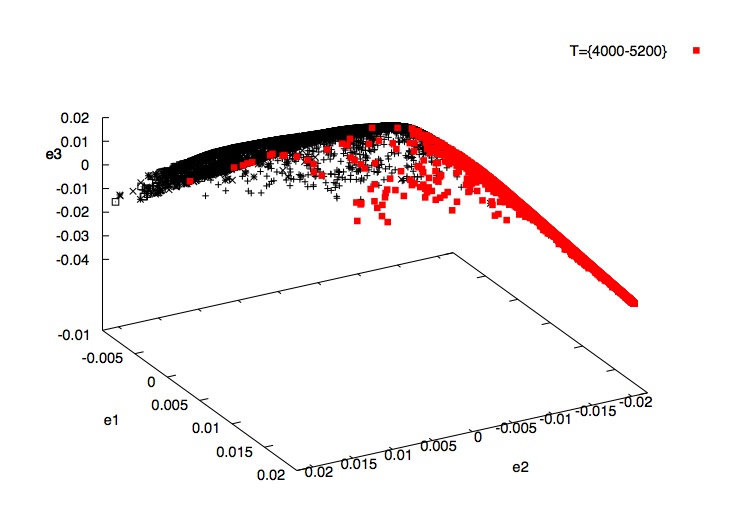}}
%{\includegraphics[scale=0.5]{boomerang_temp_just4000.eps}}
\subfigure[$5200<T_\text{eff}<6400$]
{\includegraphics[scale=0.3]{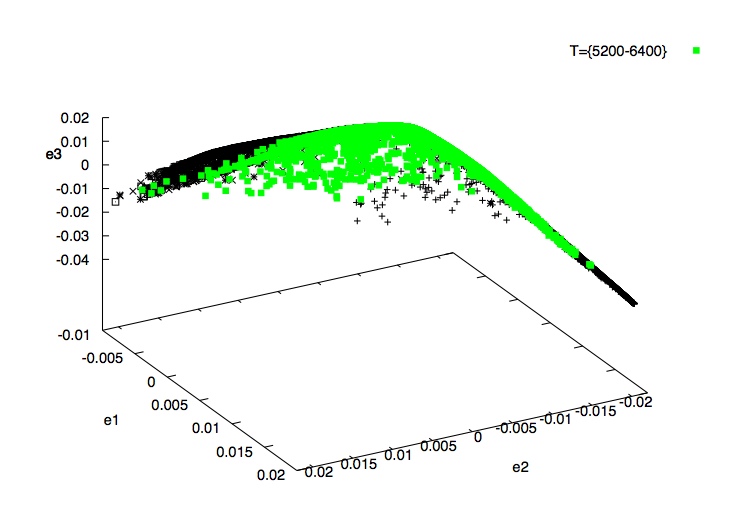}}
%{\includegraphics[scale=0.5]{boomerang_temp_just5200.eps}}
\subfigure[$6400<T_\text{eff}<7600$]
{\includegraphics[scale=0.3]{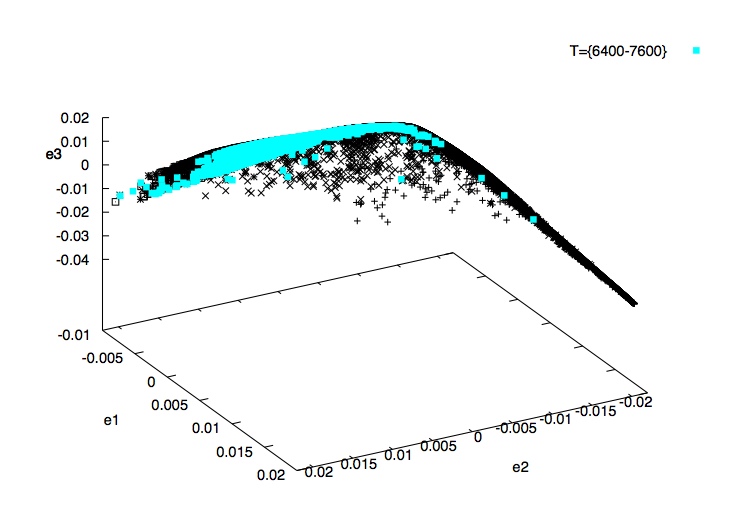}}
%{\includegraphics[scale=0.5]{boomerang_temp_just6400.eps}}
\subfigure[$7600<T_\text{eff}<8800$]
{\includegraphics[scale=0.3]{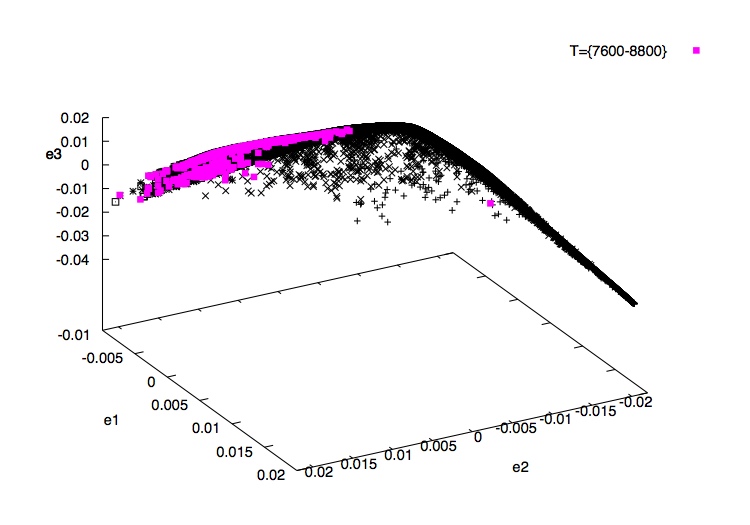}}
%{\includegraphics[scale=0.5]{boomerang_temp_just7600.eps}}
\subfigure[$8800<T_\text{eff}<10000$]
{\includegraphics[scale=0.3]{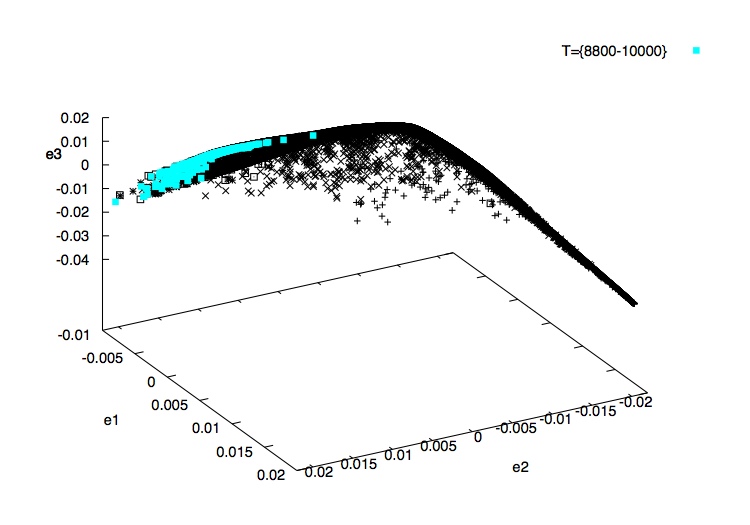}}
%{\includegraphics[scale=0.5]{boomerang_temp_just8800.eps}}
\caption{The same as Figure \ref{fig:temp} with each effective temperature
bin plotted separately to show that temperatures are not segregated in the
LLE projection space so much as they are layered.}
\label{fig:temp_layers}
\end{figure}

Given the amount of work (see Section \ref{sec:PCA}) that has gone into using
PCA to automatically classify stars according to their physical properties,
it is natural to wonder if LLE could be used to accomplish the same task.  
Figure \ref{fig:temp} plots the stellar locus of Figure \ref{fig:boom},
but this time sorts the objects by effective temperature (as determined
independently by the SDSS
sppparams\footnote{http://cas.sdss.org/dr6/en/help/docs/algorithm.asp?key=sppparams}
pipeline; data accessible from the SDSS
CasJobs\footnote{http://cas.sdss.org/CasJobs/} site).  Tracing along the
stellar locus
from left to right in that figure, one moves from high to low temperature.
Figure \ref{fig:temp_layers} reveals the details of this structure by
plotting the temperature bins from Figure \ref{fig:temp} one at a time.
While extremely high and low temperatures are confined
to specific regions in the stellar locus, 
mid-range temperatures exist throughout
the entire superstructure.
Figure \ref{fig:color} shows the same plot as Figure \ref{fig:temp} 
except that objects are now binned
according to $(g-r)_0$ color.
 
\begin{figure}[!t]
\subfigure[]
{\includegraphics[scale=0.3]{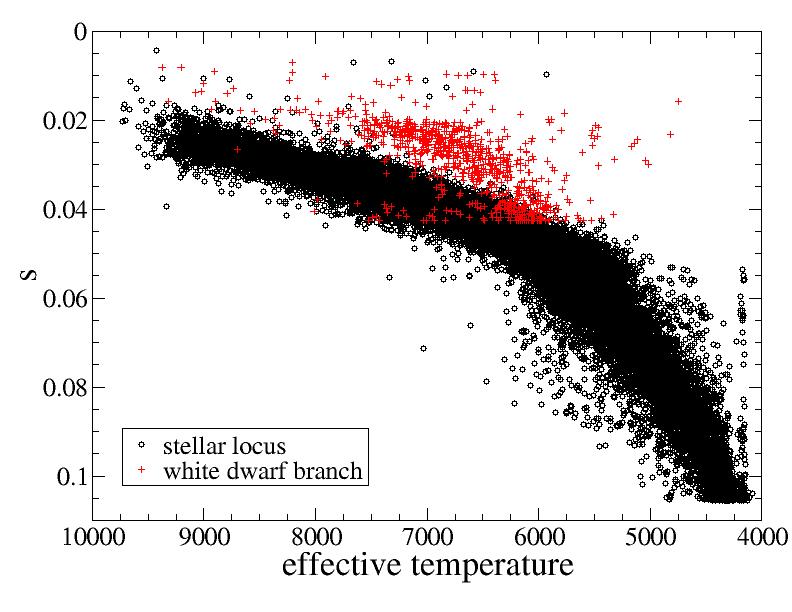}
\label{fig:t_of_s}}
\subfigure[]
{\includegraphics[scale=0.3]{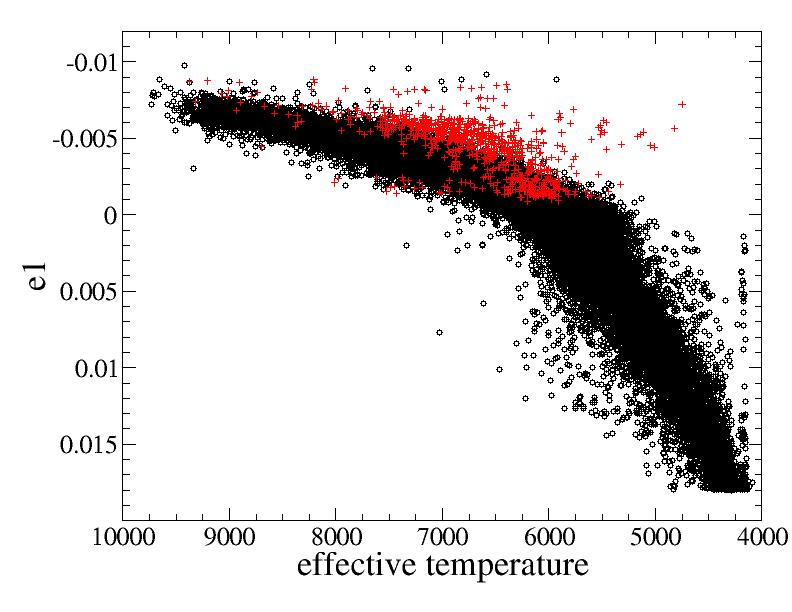}
\label{fig:t_of_x}}
\subfigure[]
{\includegraphics[scale=0.3]{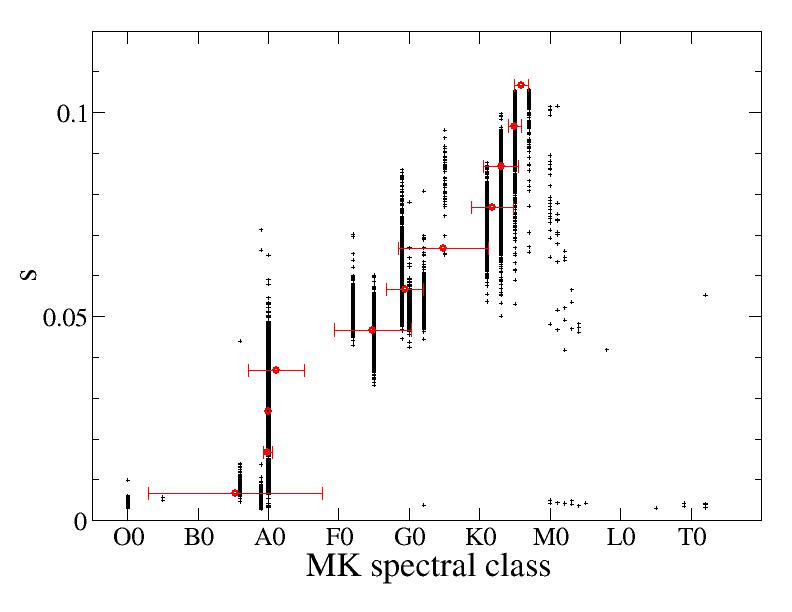}
%\includegraphcs[scale=0.3]{spec_of_s_plot.eps}
\label{fig:spec_of_s}}
\subfigure[]
{\includegraphics[scale=0.3]{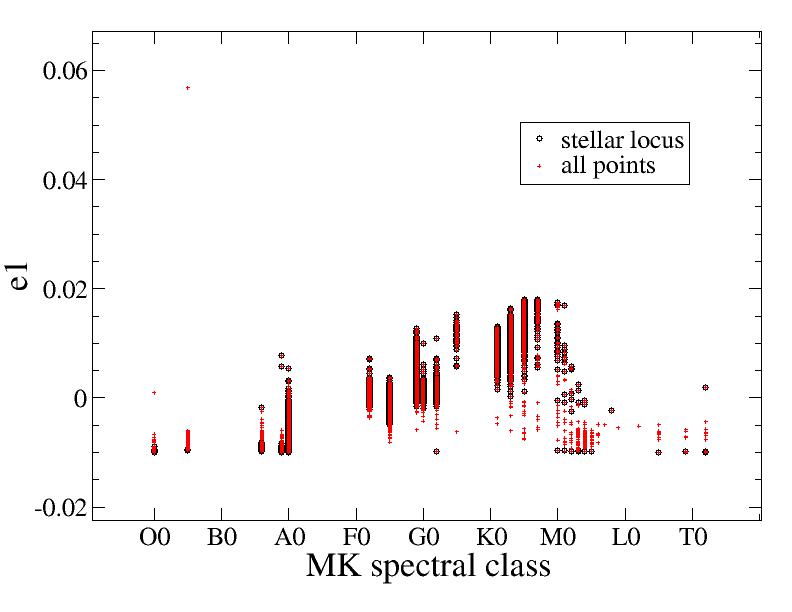}
\label{fig:spec_of_x}}
\caption{
Figure \ref{fig:t_of_s} plots effective temperatures versus $s$, the distance
along features in Figure \ref{fig:boom}.  The (black) circles are objects on the
stellar locus; $s$ is the distance along the locus.  The (red) crosses are
objects on the White Dwarf branch in Figure \ref{fig:wd}; $s$ is the distance
along the White Dwarf branch.  Figure \ref{fig:t_of_x} plots effective
temperature as a function of the coordinate $e1$ (see Figure \ref{fig:boom}) for
all objects.  Figures \ref{fig:spec_of_s} and \ref{fig:spec_of_x}
show the same plots, substituting
spectral classification for effective temperature. Crosses
(red) are all objects.  The horizontal error bars in Figure
(\ref{fig:spec_of_s}) are the result of averaging the spectral types
in bins of width $\Delta s=0.01$;
}
\label{fig:hr}
\end{figure}

\begin{figure}[!t]
\center
\includegraphics[width=\columnwidth]{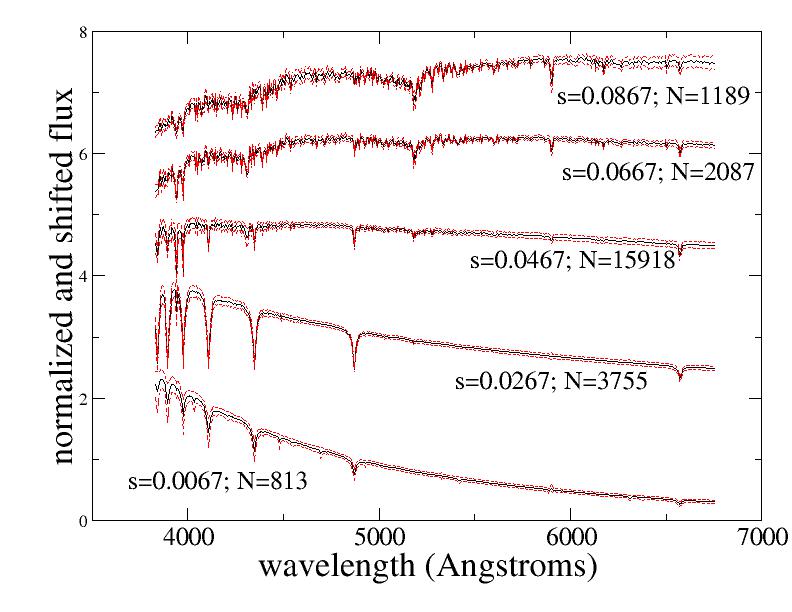}
\caption{
We plot the average spectra in bins of width $\Delta s=0.01$ where $s$
is the distance along the stellar locus in Figure \ref{fig:boom}.
Solid (black) curves indicate the mean flux at that wavelength.
Dashed (red) curves indicate the $1~\sigma$ bounds.
$N$ indicates how many spectra were averaged over.
}
\label{fig:avgspectra}
\end{figure}

\begin{figure}[!t]
\center
\includegraphics[width=\columnwidth]{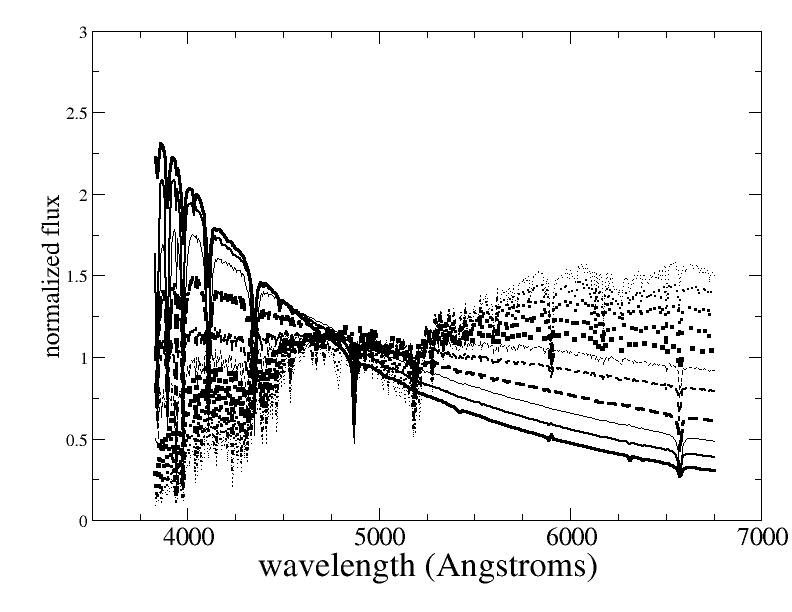}
\caption{
We plot the mean spectra in each of the $\Delta s=0.01$ bins from Figure
\ref{fig:avgspectra}, leaving out the artificial offset that separated the 
spectr in that plot.  The thickest solid curve is the bin centered on
$s=0.005$.  The thinnest dotted curve is the bin centered on $s=0.095$.
The other curves are the intervening bins in monotonic order.
}
\label{fig:meanspectra}
\end{figure}

To get a better sense of the spread in effective temperature along the stellar
locus, we want to look at the temperature of spectra as a function of their
position along the locus.  To do this, we require a quantitative means of
determining which points are ``on the stellar locus,'' which points are ``on the
White Dwarf branch,'' and which points are neither.
To this end, we parametrize both the stellar locus and the White Dwarf
branch as a series of
piece-wise $\mathcal{O}(0)$ continuous polynomials in $e1,e2,e3$.  
A point is ``on the stellar locus'' if its distance in $e1,e2,e3$ from the
parametrized curve is less than $0.001$ and is smaller than
its distance to the White Dwarf branch (this gives 43,300 of the original
49,529 spectra).  The opposite is true for objects classified as ``on the White
Dwarf Branch'' (giving 4900 spectra).
Figure \ref{fig:t_of_s} plots the effective temperature of objects 
on the stellar
locus against $s$, their
distance along the locus with $s=0$ being at the $e1=-0.0102$ end of the locus
(see Figure \ref{fig:wd}).  
The axes have been arranged and inverted to highlight the
parallels between this plot and the main sequence of the Hertzsprung-Russell
diagram. White Dwarfs (the red crosses) cluster above the ``main
sequence'' (rather than below, as in the HR diagram) when their temperatures are
plotted against distance along the White Dwarf branch.
Figure \ref{fig:avgspectra} plots the average spectra in $\Delta s=0.01$ bins
along the stellar locus.  The spectra are artificially offset to make them more
visible.  The dashed (red) curves show the $\pm 1~\sigma$ spectra about the
mean.  Figure \ref{fig:meanspectra} plots just the mean spectra in that same
binning scheme without any offset or error measurement.  
Because of the correlation between $s$ and temperature, 
the spectra pivot about a wavelength of~$\sim 4800~\AA$.  This
is reminiscent of the observation of \cite{Beauchemin:1991} that ``[the first
eigencomponent in their PCA analysis] 
weighs equally, but with an opposite sign, the spectra on each
side of\dots$4300~\AA$.''  They also found that this first 
eigencomponent correlated with temperature.  However, they needed to
include a second eigencomponent in order to accurately classify late-type stars.
This is not the case for LLE.  One dimension (either $s$ or $e1$) is adequate to
classify all types of stars, as we show below.

Figure \ref{fig:t_of_x} is effectively the same as figure \ref{fig:t_of_s}, 
except, instead of the
distance $s$ along the stellar locus, the vertical axis is just the $e1$
coordinate of the projected spectra.  The qualitative structure of the plot is
largely unchanged.  Figures \ref{fig:spec_of_s} and \ref{fig:spec_of_x} swap MK
spectral class for effective temperature and give the same plots (in this case,
black circles are stellar locus points; red crosses are all points).  In all
four cases, there is a strong, monotonic correlation between
temperature or spectral class and $s$ or $e1$.  This is especially interesting
given the findings of previous works that spectral classification using PCA
requires at least two eigenspectra to attain any reasonable accuracy
\cite{Singh:1998,Singh:2001,StorrieLombardi:1994,Whitney:1983a,Whitney:1983b,Christian:1982}.
In the words of Singh {\it et al}. (2001), this is because ``Spectral type has a
complex dependence on spectral features, whereas the [Principal Components] are
just linearly related to the original spectra'' (see especially their Figure 6).
By accounting for the non-linear relationships between neighborhoods of spectra,
LLE seems to have overcome this hurdle.

%This also lends credence to our hypothesis that LLE will prove 
%an efficient way to characterize stellar spectra.

%uses sdss_stars_gnu_temp_2.sav

%%%%%%%%%%%%%%%%%%%%%%%%%%%%%%%%%%
\section{Discussion and Conclusions \label{sec:discussion}}

The results of section \ref{sec:results} demonstrate the efficacy of LLE as an
automatic stellar classification algorithm.  In the sections below, we will
directly compare LLE to PCA-based methods and propose a more detailed
implementation of LLE for future data sets.

\subsection{Comparison to PCA \label{sec:compare}}

Figures \ref{fig:boom} and \ref{fig:wd} of this work 
show the efficacy of LLE at separating
spectra by the physical nature (e.g. galaxy, variable star, white dwarf, main
sequence star, etc.) of the emitting object.  Cabanac {\it et al.} (2002)
demonstrated a similar utility of PCA however, comparing our figures with their
Figure 15, we see that LLE provides a much more straightforward and identifiable
division of objects.

From Figures \ref{fig:hr} of this work, we see that LLE is also useful for
separating out more detailed physical characteristics (i.e. effective
temperature and spectral classification) of objects.  This has been a major goal
of PCA work for several decades.
Figure 3(d) of Storrie-Lombardi {\it et al}. (1994) shows the efficacy of PCA
projection fed into a neural network as a means of accurately reproducing the
MK classification of stellar spectra.  
Their automated classification is arguably
more accurate than what you might infer from our Figures \ref{fig:hr}.
However, they require five eigencomponents to achieve that accuracy.  
Conversely, the correlation
between MK classification and LLE projection is largely one dimensional. 
This is especially evident from our Figure \ref{fig:spec_of_x}, 
which shows that,
without any additional processing, the $e1$ LLE dimension monotonically maps
onto MK spectral classification. 
This mapping is accurate to within a few spectral sub-types.
This is the same accuracy achieved with PCA.  Recall that
we search specifically for a one-dimensional classification scheme.
The simplicity of this objective directly limits the accuracy with which
we can hope to classify our objects.  It is an improvement over
the more complex schemes derived from PCA.
Christian (1982) attempted to directly reproduce the MK classification using
Principal Components.  Results from that paper showed that, 
while the first eigencomponent correlated
with spectral type, the correlation was nonlinear, requiring the data set to be
divided into early- and late-type stars, each set being analyzed separately. 
Such nonlinearity is not apparent in our Figure \ref{fig:spec_of_x}.  LLE
is robust against nonlinearities in the data.

McGurk {\it et al}. (2010) showed a correlation between PCA projection
and the metallicity of stars.  We attempted a similar analysis and found no
obvious correlation between LLE dimensions and metallicity.  

\subsection{Recursive LLE \label{sec:layers}}

\begin{figure}[!t]
%\subfigure[]{\includegraphics[scale=0.6]{nocarbon_temp_jpg.eps}
\subfigure[]{\includegraphics[scale=0.85]{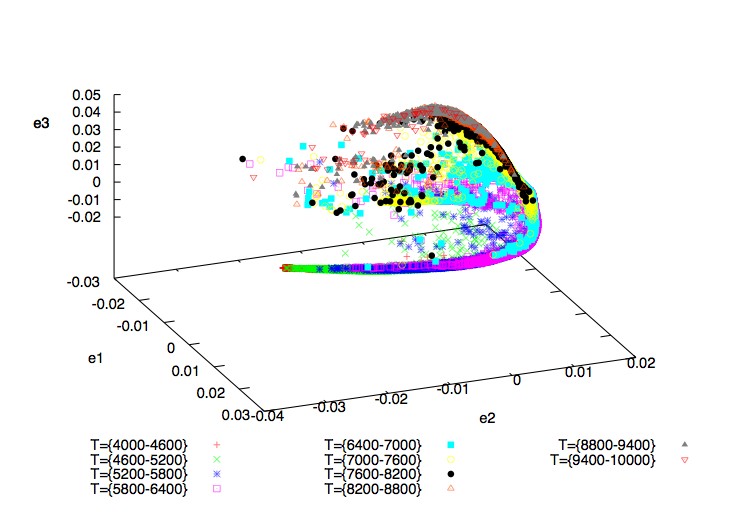}
\label{fig:nocarbontemp}}
%\\\subfigure[]{\includegraphics[scale=0.6]{nocarbon_color_jpg.eps}
\\\subfigure[]{\includegraphics[scale=0.85]{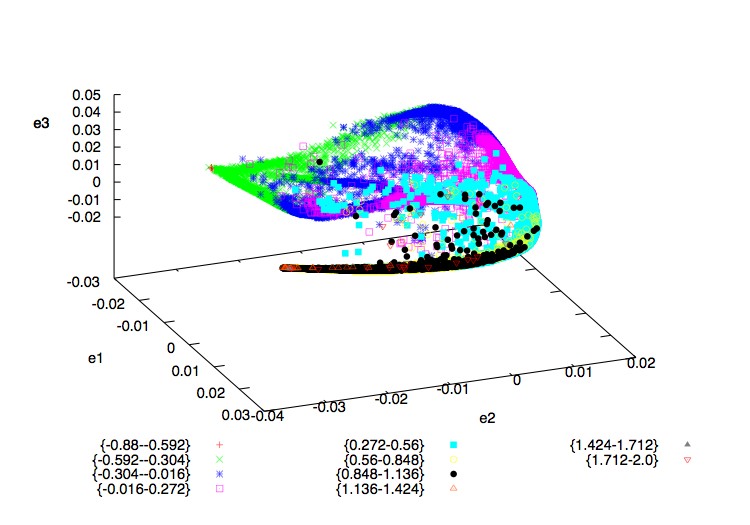}
\label{fig:nocarboncolor}}
\caption{
The LLE projection of our data with the galaxy and CV branches from Figure
\ref{fig:boom} and the carbon star branch from Figure \ref{fig:justboom}
excised.  Figure \ref{fig:nocarbontemp} color codes the data points according to
effective temperature.  Figure \ref{fig:nocarboncolor} color codes the data
points according to $(g-r)_0$ color.
}
\label{fig:nocarbon}%
\end{figure}

The progression from Figures \ref{fig:boom} to Figure
\ref{fig:justboom} suggests a hierarchical system of layered LLE
projections in which each successive projection discards the extreme
outliers from its immediate predecessor.  That is to say, in the same
way that we discarded the galaxy and CV branches identified in Figures
\ref{fig:boom} to generate Figure \ref{fig:justboom}, we can perform
another projection, this time additionally discarding those objects
that lie in the carbon star branch ($e2<0$, $e3\rightarrow\infty$) in
Figure \ref{fig:justboom}.  Figures \ref{fig:nocarbon} plot this
projection by both effective temperature and color.  In this
projection, the stellar locus (recall that we have now removed all of
the points that do not lie on the stellar locus in Figure
\ref{fig:boom}) is no longer one-dimensional, and, while the
progression in temperature (Figure \ref{fig:nocarbontemp}) first
observed in Figure \ref{fig:temp} remains, it is not nearly as clean
as the progression in color (Figure \ref{fig:nocarboncolor}).  By
removing the outlier populations, we have allowed more subtle
relationships between temperature and color to manifest themselves in
the LLE projection.  One can
imagine continuing this process indefinitely, refining the information
gleaned from LLE with each new projection.

%%%%%%%%%%%%%%%%%%%%%%%%%%%%%
\acknowledgments
SFD and AJC acknowledge support from DOE award grant number DESC0002607.
AJC would like to thank Andrew Hopkins and the Australian Astronomical
Observatory for their support of this work and their wonderful hospitality
during his visit to the AAO as part of the Distinguished Visitor Program.

%%%%%%%%%%%%%%%%%%%%%%%%%%%%%%%%%%%%%%%%%%%%%%%%%%%%%%%%%%%%
%%%%%%%%%%%%%%%%%%%%%%%%%%%%%%%%%%%%%%%%%%%%%%%%%%%%%%%%%%%%

\end{document}